\documentclass[final]{article}
\usepackage{graphics}
\usepackage{graphicx}

\newcommand{\be}{\begin{equation}}
\newcommand{\ee}[1]{\label{#1} \end{equation}}
\newcommand{\ba}{\begin{eqnarray}}
\newcommand{\ea}[1]{\label{#1} \end{eqnarray}}
\newcommand{\nl}{\nonumber \\}

\newcommand{\etal}{ et.al. }

\begin{document}

\title{{\bf Abstract composition rule for relativistic kinetic energy in the thermodynamical limit}}

\author{ 
 \centerline{ T. S. Bir\'o} \\[1em]
  {KFKI Research Institute for Particle and Nuclear Physics,} \\
  { H-1525 Budapest, P.O.Box 49, Hungary} 
}


\date{}
\maketitle

\abstract{
 We demonstrate by simple mathematical considerations that 
 a power-law tailed distribution in the kinetic energy of
 relativistic particles can be a limiting distribution seen in relativistic heavy ion experiments.
 We prove that the infinite repetition of an arbitrary composition rule on an infinitesimal
 amount leads to a rule with a formal logarithm. As a consequence the stationary distribution
 of energy in the thermodynamical limit follows the composed function of the Boltzmann-Gibbs
 exponential with this formal logarithm. In particular, interactions described as solely functions
 of the relative four-momentum squared lead to kinetic energy distributions
 of the Tsallis-Pareto (cut power-law) form in the high energy limit.
}

\normalsize


\section{Why to investigate composition rules}

One of the theoretically challenging questions related to relativistic heavy ion physics
is how to establish the existence of a thermal or near thermal state of (quark) matter
during high energy collisions. Besides studies of the hadronic flavor composition
\cite{SM1,SM2,SM3,SM4,SM5}, 
the shape and the steepness of transverse momentum spectra
on different particles presents experimental information on this 
question\cite{PHENIX1,PHENIX2,PHENIX3,PHENIX4,STAR1,STAR2,STAR3}.
Since for high energies of the out-coming particles (pions, kaons, antiprotons, etc.)
these spectra show a power-law tail, it is important to clarify, whether this prominent 
feature can be explained in the framework of general thermodynamic ideas, or it has
to be rendered to the realm of case by case non-equilibrium dynamics.
While in some earlier works we have demonstrated that such spectra can be obtained as
stationary distributions by altering the energy composition rule in two-particle collisions
from the simple addition to another rule \cite{NEBE,PLAnoneq}, in the present paper we
aim at understanding the general mechanism setting an effective rule and a non-exponential
stationary distribution of the individual particle energies in the thermodynamical limit.  

The classical thermodynamics using extensive and intensive quantities
is being extremely successful in describing and
understanding uncountable physical phenomena in nature. However, there 
are particular cases, where the observed distribution of individual energies
in a complex system does not follow the exponential Boltzmann-Gibbs law.
There are suggestions trying to go beyond the classical picture and to consider descriptions
generalizing traditional thermodynamical concepts, in particular to consider the possibility
that the composition rule deviates from the simple addition. Such quantities are often
called 'non-extensive', although strictly speaking the extensivity property is required
only in the thermodynamical limit, i.e. for large systems consisting of many particles.

In particular generalizations of the entropy formula, connecting the quantity of macroscopic 
entropy to probabilities of microstates of an extended system, have been repeatedly suggested
in forms generalizing the Boltzmann-Gibbs-Shannon logarithmic formula\cite{GE1,GE2,GE3,GE4}. 
The deformed (or generalized,
extended) logarithm relates the abstract product rule to a summation or composition formula
in a general way: the statistical independence of states is hence mapped to a non-additivity
of the composition formula\cite{GE5,GE6,GE7,GE8,GE9}. 
Reversely, the additivity of entropy is - in some cases - 
achieved by non-product probabilities, trying to grasp the essence of surviving correlations 
(surmised to occur due to long range interactions) in systems, which are large in the 
thermodynamical sense\cite{ADDITIVE_NOT_LOG_ENTROPY}. 
The additivity of entropy also can be achieved by weakly non-local extensions \cite{Fis59b,Van06a}.
Naturally, the use of a formal composition rule is also applicable to the energy\cite{NON_ADD_ENERGY,WANG}. 
Two subsystems combine to a common larger system not necessarily additively with their energies;
the interaction part may lead to a finite relative contribution in the infinite particle
number limit. 


\section{Repeated compositions and the proof of associativity}

The important question arises, that what happens if we repeat (compose with itself) an abstract
mathematical composition rule (the formal product in group theory) infinitely many times, but
each time applying to an infinitesimal amount: grasping this way the extensivity limit of
arbitrary composition rules. 
We find that some rules transform to the simple addition in this limit, while others not.
The proposition is that the thermodynamical limit of an arbitrary pairwise, iterable
composition rule is an {\em associative} rule. The importance of this statement becomes clear
by referring to the known mathematical property that associative rules always possess
a strict monotonic function, called here the formal logarithm, in terms of which they can be 
expressed\cite{MATH_OF_FORMAL_LOG}.
Let us denote an abstract pairwise composition rule by the mapping 
$(x,y) \rightarrow h(x,y)$. Whenever $h(x,y)$ is an element of the same set as $x$ and $y$,
the composition is iterable arbitrarily long.

Associativity of such a rule is formulated by the function equation
\be
 h(h(x,y),z) \: = \: h(x,h(y,z))
\ee{ASSOC}
for $x,y$ and $z$ being elements of the same group. For our purpose we shall consider energies or 
entropies of physical subsystems.
The general solution of the associativity equation (\ref{ASSOC}) is given by
\be
 h(x,y) = X^{-1}\left( X(x)+X(y)\right)
\ee{FORMLOG}
with  $X(x)$ being a strict monotonic function. It is referred to as the ''formal logarithm'',
because it maps the arbitrary composition rule $h(x,y)$ to the addition by taking the 
$X$-function of eq.(\ref{FORMLOG}):
\be
 X(h(x,y)) = X(x) + X(y).
\ee{FORMADD}
Due to this construction the generalized analogs to classical extensive (and additive) quantities
are formal logarithms, whenever the composition rule is associative. As a consequence stationary
distributions, in particular by solving generalized Boltzmann equations \cite{NEBE}, are the Gibbs
exponentials of the formal logarithm,
\be
 f(x) = \frac{1}{Z} e^{-\beta X(x)}.
\ee{GIBBSFORM}

Now we investigate a large number of iterations, $N$, of the composition rule applied to
an infinitesimal amount $y/N$ in each step:
\be
 x_N(y) := \underbrace{h \circ\ldots \circ h}_{N-1} \left(\frac{y}{N}, \ldots, \frac{y}{N}\right). 
\ee{ITER} 
Whenever the limit, 
\be
 \lim_{N\rightarrow\infty}\limits x_N(y) < \infty,
\ee{LIM}
is finite for a finite $y$,
we are dealing with an extensive (but not necessarily additive) system.
Our purpose is to study such systems and to obtain their asymptotic composition rule,
\be
 x_{N_1+N_2} = \varphi(x_{N_1},x_{N_2})
\ee{ASYMPCOMP}
in the limit $N_1, N_2 \rightarrow \infty$.
The repetitive composition can be formulated as a recursion at an arbitrary step $n$ between
$0$ and $N$ as follows:
\be
 x_n = h\left(x_{n-1},\frac{y}{N}\right),
\ee{RECURR}
with $x_0=0$. 
It is a natural requirement, but important for that what follows, to consider only such
rules which satisfy $h(x,0)=x$.
Subtracting $x_{n-1}=h(x_{n-1},0)$ from this formula we arrive at
\be
 {x_n-x_{n-1}} = {h(x_{n-1},\frac{y}{N})-h(x_{n-1},0)}.
\ee{DISCRETE}
Introducing the (in the thermodynamical limit continuous) variable $t=n\epsilon$,
we follow an evolution in $t$ alike the renormalization flow:
\be
 \frac{dx}{dt} = \frac{y}{t_f} \: h_2'(x,0^+).
\ee{FLOW}
In the expression on the right hand side
$h_2'(x,0^+)$ denotes the partial derivative of the rule $h(x,y)$ with respect to its
second argument taken at this argument value approaching zero from above. The final time
is given by $t_f=N\epsilon$. Note that the uniformity of subdivisions to $y/N$ is
not really necessary; all infinitesimal divisions summing up to $t_f$ lead to the same
differential flow equation.

The solution of eq.(\ref{FLOW}) is given by
\be
 L(x) = \int_0^x\limits \frac{dz}{h_2'(z,0^+)} \, = \, y \, \frac{t}{t_f}.
\ee{ASYLOG}
This solution, when strict monotonic and hence invertible, defines the following asymptotic
composition rule in the thermodynamic limit:
\be
 x_{12} := \varphi(x_1,x_2) = L^{-1} \left( L(x_1)+L(x_2) \right),
\ee{ASYCOM}
due to $L(x_1)=yt_1/t_f$, $L(x_2)=yt_2/t_f$ and $L(x_{12})=y(t_1+t_2)/t_f$.
Note that $t_i/t_f = N_i/N$ are the extensivity shares of the respective subsystems.
This asymptotic composition rule is associative and commutative.


\section{Classification, important examples}

Now we turn to the analysis of important particular rules and their asymptotic
pendants in the thermodynamic limit.

The trivial (and classical) addition is the simplest composition rule: $h(x,y)=x+y$.
In this case $h_2'(x,0^+)=1$ and one obtains
\be
 L(x) = \int_0^x\limits dz = x,
\ee{ADDFORMALLOG}
and with that the original Gibbs exponentials, $e^{-\beta E}/Z$, for stationary
distributions of any Monte Carlo type algorithm using the original composition rule. 
The corresponding asymptotic rule is  $\varphi(x,y)=x+y$.

The rule leading to the $q$-exponential\cite{GE5} (or Pareto, or Tsallis) distribution is given by
$h(x,y)=x+y+axy$ with the parameter $a$ proportional to $q-1$. In this case one obtains
$h_2'(x,0^+)=1+ax$ and
\be
 L(x) = \int_0^x\limits \frac{dz}{1+az} = \frac{1}{a} \ln(1+ax).
\ee{TSALLISLOG}
This formal logarithm leads to a stationary distribution with power-law tail
as the function composition $exp \circ L$ on the power $-\beta$:
\be
 f(E) = \frac{1}{Z} e^{- \frac{\beta}{a} \ln(1+a E)} = \frac{1}{Z} \left( 1+a E\right)^{-\beta/a}.
\ee{TSALLISEXP}
A generalized entropy formula on the other hand can be constructed
as the expectation value of the inverse of this function,
$L^{-1} \circ \ln$:
\be
 S = \int\! f \, \frac{e^{-a\ln(f)}-1}{a} \: = \: \frac{1}{a} \int  \, (f^{1-a}-f).
\ee{TSALLIS_ENTROPY}
The asymptotic composition rule again coincides with the original one:
$\varphi(x,y)=x+y+axy$.

A simple rule suggested by Kaniadakis\cite{KANIADAKIS} 
is based on the $\sinh$ function. The formal logarithm is given as
\be
 L(x) = \frac{1}{\kappa} {\rm Ar sh} (\kappa x),
\ee{KANI_LOG}
and its inverse becomes $L^{-1}(t)=\sinh(\kappa t)/\kappa$.
The stationary distribution, composed by \hbox{$exp \circ L$}, is
\be
 f_{{\rm eq}}(p) = \frac{1}{Z} \left(\kappa p + \sqrt{1+\kappa^2p^2}\right)^{-\beta/\kappa}.
\ee{KANI_DIST}
For large arguments it gives a power-law in the momentum absolute value $p$ and hence also in the
relativistic energy.
The corresponding entropy formula is the average of $L^{-1}\circ \ln$ over the allowed phase space:
\be
 S_K = - \int \frac{f}{\kappa} \sinh(\kappa\ln f) = \int \frac{f^{1-\kappa}-f^{1+\kappa}}{2\kappa}.
\ee{KANI_ENT}
The composition formula can be reduced to
\be
 h(x,y) = x\sqrt{1+\kappa^2y^2} + y\sqrt{1+\kappa^2x^2}.
\ee{KANI_ADD}
For low arguments it is additive, $h(x,y) \approx x+y$, for high ones it is multiplicative,
$h(x,y) \approx 2\kappa xy$. It has been motivated by the relativistic kinematics of
massive particles. 
Regarding $\kappa  = 1/mc$, $\kappa p = \sinh \eta$, 
so the formal logarithm is proportional to the rapidity, $L(p)=mc\eta$.
This implies a stationary distribution like $exp(-\beta mc|\eta|)$, which has not
yet been observed in relativistic particle systems.
Therefore it is wishful to consider some further scenarios based on other quantities
deduced from relativistic kinematics (see next section).

The rule leading to a stretched exponential stationary distribution is given by
$h(x,y)=\left(x^b+y^b\right)^{1/b}$. Here some care has to be taken, the partial derivative
has to be evaluated not at zero, but at a small positive argument, $\epsilon=y/2N$.
We get $h_2'(x,\epsilon)=c(\epsilon)x^{1-b}$ with a factor depending on $\epsilon$
and - depending on $b$ - possibly diverging in the $\epsilon=0$ limit. However, this
does not spoil our procedure; we obtain $L(x)=c(\epsilon)x^b/b$ and with that
the asymptotic rule:  $\varphi(x,y)=\left(x^b+y^b\right)^{1/b}$. The reason is that constant
factors in the formal logarithm can be eliminated without loss of any information.

Our next example is a non-associative rule; its asymptotic pendant cannot be itself.
We consider
\be
 h(x,y) = x + y + a \frac{xy}{x+y}
\ee{NONASSOC}
(a combination of arithmetic and harmonic means). The fiducial derivative is given by
$h_2'(x,0^+)=1+a$ and -- being a constant -- it leads to $L(x)=x/(1+a)$ and
with that to the {\em addition} as asymptotic rule: $\varphi(x,y)=x+y$.

As a last example in this train we discuss Einstein's formula for velocity addition,
\be
 h(x,y) = \frac{x+y}{1+xy/c^2}.
\ee{EINSTEIN}
This rule is associative, and it also preserves its form in the thermodynamic limit.
The fiducial derivative is given by $h_2'(x,0^+)=1-x^2/c^2$ and the formal logarithm,
$L(x)= \, c {\rm \, atanh\: } (x/c)$ turns out to be the rapidity. The asymptotic composition
rule recovers the original one.

In the case of a general second order polynomial for the fiducial derivative $h_2'(z,0)$ the
asymptotic composition rule turns out to be:
\be
 \varphi(x,y) = \frac{x+y+axy}{1+xy/c^2}
\ee{SECONDORDER}
with $c^2=-z_1z_2$ and $a=-(z_1+z_2)/z_1z_2$;  
$z_1$ and $z_2$ being the algebraic roots of $h_2'(z,0)$. 
It is a generalization of the Tsallis and
Einstein rules. A similar composition rule has been found for the parallel transmittivity in
certain Potts models, for a review see  Ref.\cite{TSALLIS_REPORT}.

Finally we prove that {\em all associative rules are mapped to themselves by the thermodynamic limit}.
Given an original composition rule, $h(x,y)$, which is associative, it can be expressed
by its formal logarithm:
\be
 h(x,y) = X^{-1}\left(X(x)+X(y) \right).
\ee{STRATASSOC}
Then letting to act the strict monotonic function $X$ on both sides and derive with
respect to the second argument we obtain $X'(h)\, \partial h/\partial y = X'(y)$ and
\be
 h_2'(x,0^+) = \frac{X'(0)}{X'(h(x,0))}.
\ee{FIDUCIAL_DERIV2} 
Due to the property $h(x,0)=x$ (equivalently $X(0)=0$)
the formal logarithm of the asymptotic composition rule is given by
\be
 L(x) = \int_0^x\limits \frac{X'(z)}{X'(0)}dz = \frac{X(x)}{X'(0)};
\ee{FORMAL_ASYMP}
it is proportional to the formal logarithm of the starting rule. Therefore
the asymptotic rule is the same as we begun with: $\varphi(x,y)=h(x,y)$.
Actually the freedom in a factor of the formal logarithm always can be used to set $X'(0)=1$.

This way any associative composition rule describes a thermodynamical limit of a class of
non-associative rules. Associativity is synonym to the thermodynamical limit.

\begin{figure}

\begin{center}
\includegraphics[width=0.60\textwidth,angle=-90]{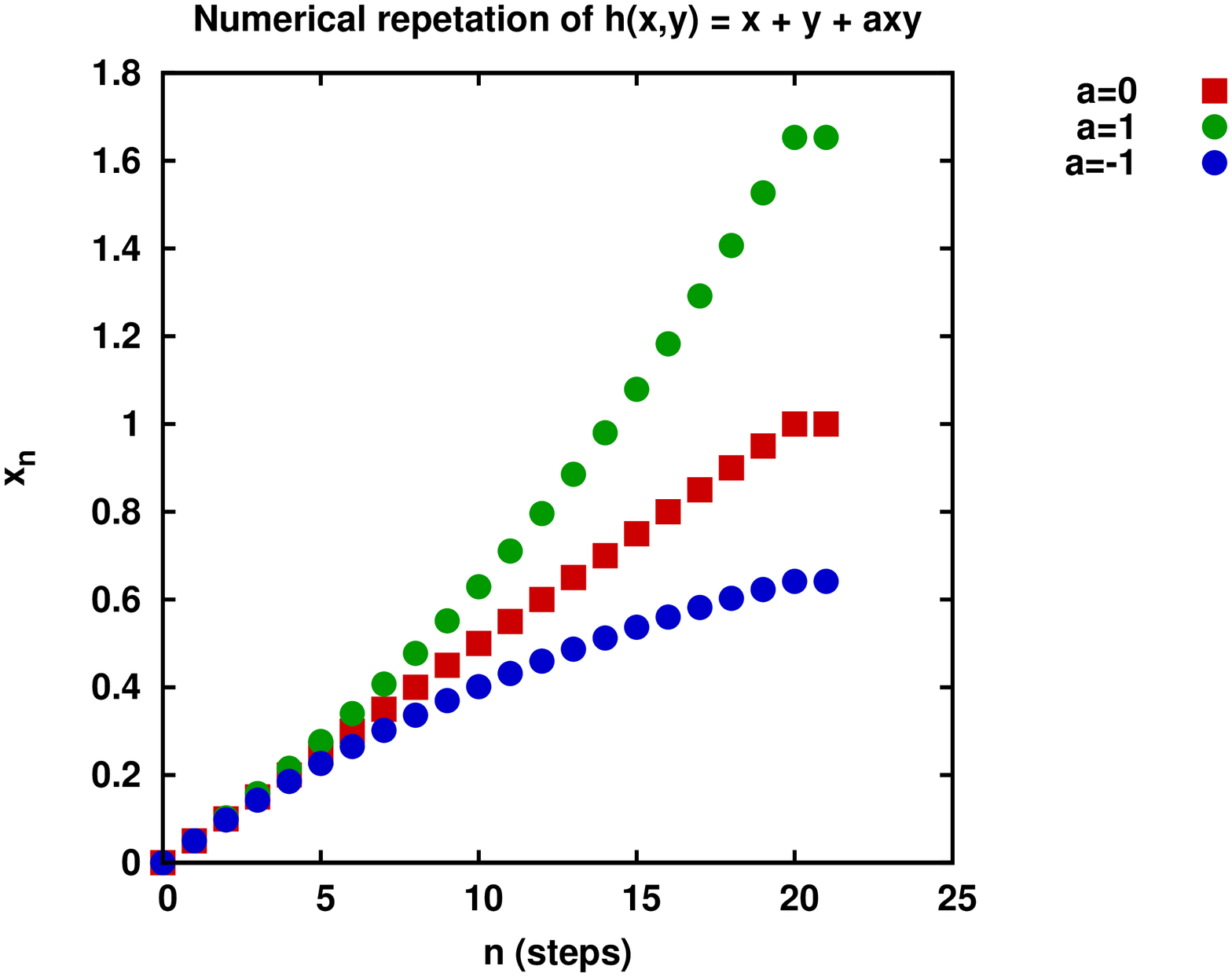} 
\includegraphics[width=0.60\textwidth,angle=-90]{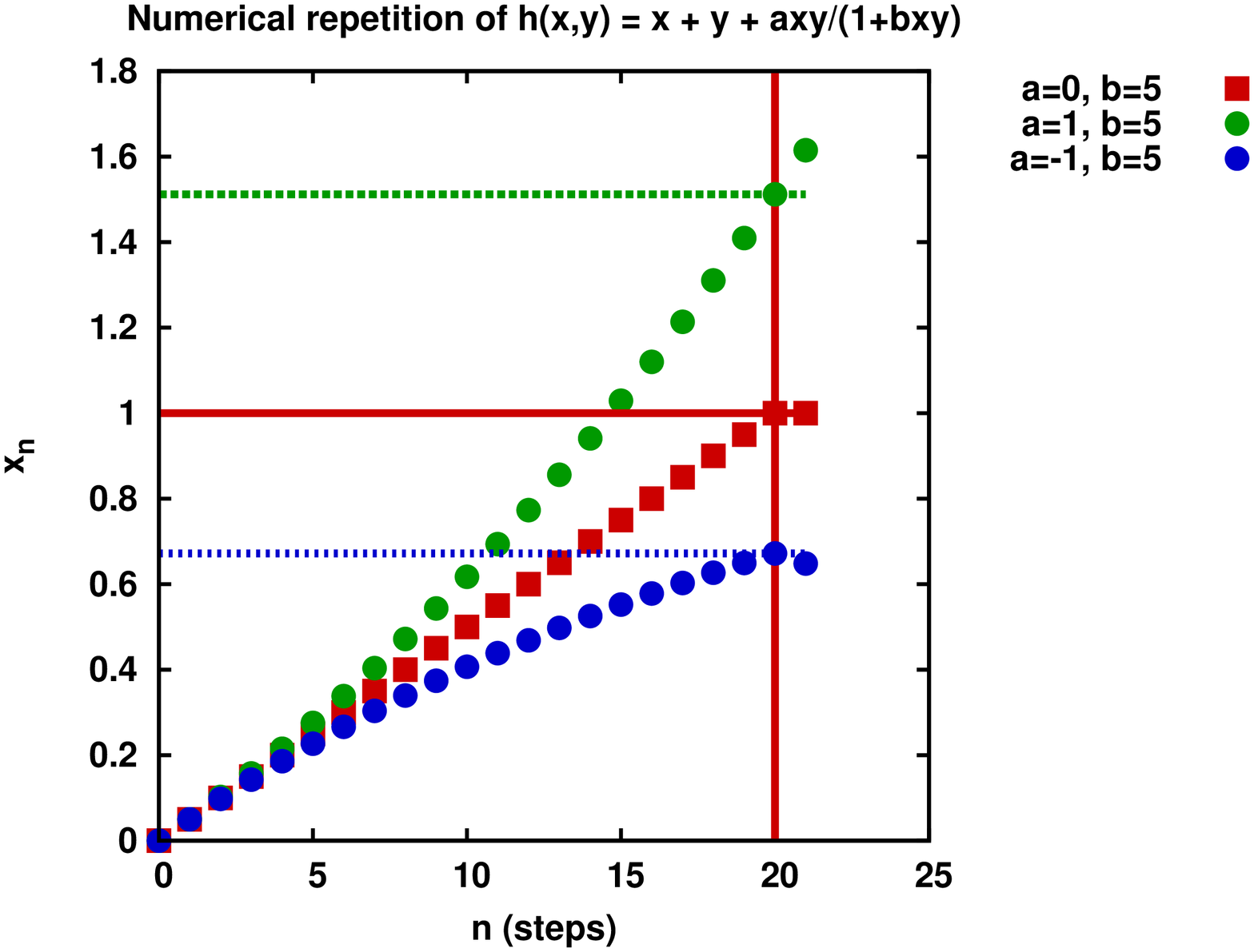} 
\end{center}
\caption{ \label{Fig1}
 The series of composed amounts $x_N$ by subdividing $y=1$ to $N$ parts.
 In the top figure (a) results with the asymptotic rule $h(x,y)=x+y+axy$,
 while in the bottom (b) with a non-associative rule $h(x,y)=x+y+xy/(1+5xy)$.
 The 21-st values are the asymptotic compositions of half systems: $\varphi(x_{10},x_{10})$.
}

\end{figure}

In Fig.\ref{Fig1} we show the composition of $y=1$ from $y/2N$-sized pieces by the rule
$h(x,y)=x+y+G(xy)$ for $G(w)=aw/(1+bw)$ with different parameters $a$ and $b$.
While the $b=0$ case represents the Tsallis-Pareto rule (top figure, a), which asymptotically establishes,
the $b=5$ case plot deviations from this for finite $N$-s (bottom figure, b).
The 21-st point is the asymptotic composition of half-sized systems, $\varphi(x_N,x_N)$,
the other points plot $x_n$ by the recursive application of the rule. 
One inspects some deviation for $b \ne 0$, while for an associative rule $h(x,y)$ no deviation
in the end result occurs.


\begin{figure}
\centerline{\includegraphics[width=0.60\textwidth,angle=-90]{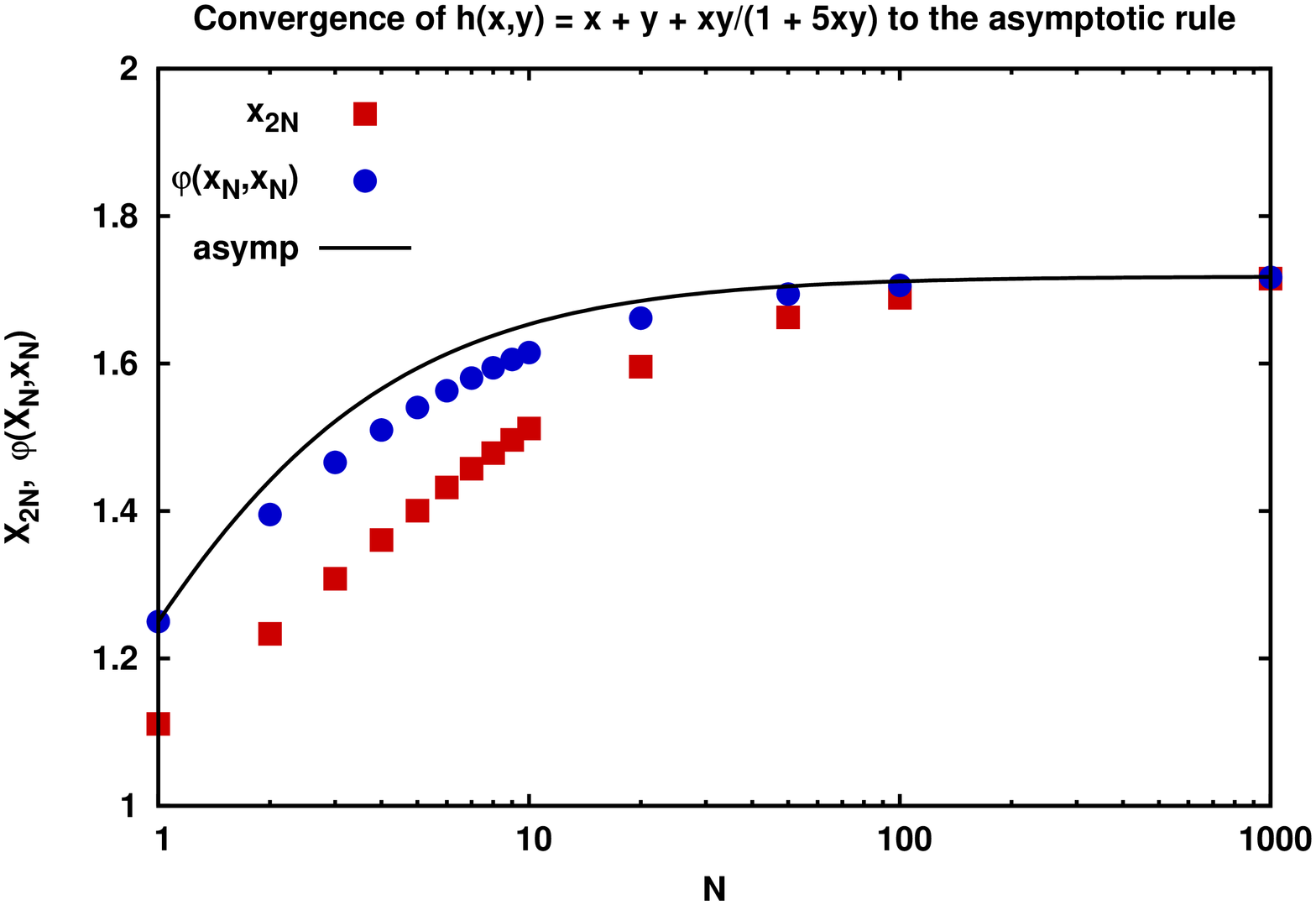}}
\caption{\label{Fig2}
 Approach to the asymptotic extensivity rule as a function of the number of repetitions 
 of the non-associative rule, $h(x,y)=x+y+xy/(1+5xy)$, to the amount $1/N$ (boxes). Plotted are
 $x_{2N} $ and for comparison the amount $\varphi(x_N,x_N)$ as well as the curve corresponding
 to the $2N$-fold composition of $\varphi$. 
}
\end{figure}

Fig.\ref{Fig2} shows the approach to the asymptotic limit as a function of the
number of repetitions, $N$ of the micro-rule, $h(x,y)$ on a logarithmic scale.
The series of $x_{2N}$ values is constructed by using $h(x,y)=x+y+xy/(1+5xy)$ --
which is not an associative rule. The asymptotic rule corresponding to this choice is
given by $\varphi(x,y)=x+y+xy$. The higher points belong to $\varphi(x_N,x_N)$, they 
deviate from $x_{2N}$ for finite N. The continuous curve represents the 
$2N$-fold composition of $1/2N$ according to the asymptotic rule $\varphi(x,y)$.
It is given by the formula
\be
 x_{2N} = \underbrace{\varphi\circ\ldots\circ\varphi}_{2N-1} \, \left(\frac{1}{2N}, \ldots, \frac{1}{2N}\right)
        = \left(1+\frac{1}{2N} \right)^{2N} - 1.
\ee{FORMULA}
The asymptotic rule is occurring slowly in this case, the convergence rate is given by
the convergence of the Euler formula to the Euler number ($x_{\infty}=e-1$).

\section{Deriving composition rules from interaction energy and kinematics}

In this section we review a few general considerations which may relate the 
interaction energy due to some kind of correlation to the use of formal logarithms and abstract
composition rules. Our basic assumption is that the interaction energy between two subsystems
('particles') can be expressed as a function of the individual energies without the interaction
(in asymptotic free states). This way $E_{12}=E_1+E_2+U(E_1,E_2)$ is a general energy composition rule.

In most cases, discussed in physics, the interaction is given as a function of the relative
distance, and this form is not directly related to kinematic data. In the sense of medium long
time behavior, however, the interaction energy can often be expressed by the relative momentum:
either due to a virial theorem or due to direct quantum mechanical solution for the relative 
wave function, like e.g. in the two-body Coulomb problem. 

In the followings we shall assume that the interaction energy is a function of the
kinematic variable $Q^2$, the square of relative four-momentum. We shall study whether relativistic speeds
alone can cause ''non-extensivity'', i.e. a power-law tailed kinetic energy distribution. 
The relativistic formula for $Q^2$ is given by the following Lorentz-invariant quantity:
\be
 Q^2 = (\vec{p}_1-\vec{p}_2)^2 - (E_1-E_2)^2
\ee{Q2_LORENTZ}
with $\vec{p}_i,E_i$ being relativistic momenta and full energies of interacting bodies.
Expressed by the energies and the angle $\Theta$ between the two momenta this becomes
a linear expression of $\cos\Theta$:
\be
 Q^2 = 2 \left(E_1E_2-m^2 - p_1p_2\cos\Theta   \right)
\ee{Q2_THETA}
with $p_i=\sqrt{E_i^2-m^2}$ for $i=1,2$. Here we use relativistic units ($c=1$) and assume the same
mass for both interacting partners, for simplicity. It is useful to note that writing eq.(\ref{Q2_THETA})
as $Q^2=2(A-B\cos\Theta)$ we have
\be
 A \pm B = E_1E_2-m^2 \pm p_1p_2 = m^2 \left( \cosh(\eta_1 \pm \eta_2) - 1 \right)
\ee{AB}
with using the rapidities $\eta_i$. For equal momenta or rapidities
$Q^2=0$ and $A=B$;  $A^2-B^2=m^2(E_1-E_2)^2$ measures the energy difference.

In order to estimate the interaction contribution we subtract the zero momentum terms,
and assume
\be
 E_{12} = E_1 + E_2 + U(Q^2) - U(Q_1^2) - U(Q_2^2) + U(0),
\ee{TOTAL_E12}
with $Q_i^2=2m(E_i-m)$. This construction ensures that for equal momenta,
i.e. for $Q^2=0$, no interaction correction occurs to the addition law for the energy.

Seeking for an effective energy composition rule as an isotropic average over the relative
directions of the respective momenta, one averages over the angle $\Theta$:
\ba
 \langle U(Q^2) \rangle &=& \frac{1}{2} \int_{0}^{\pi}\limits U(2A-2B\cos\Theta) \sin\Theta \, d\Theta
\nl
& =& \frac{F(2A+2B)-F(2A-2B)}{4B}, 
\ea{U_ISO}
with $U(w)=dF/dw$. This is easy to see upon the substitution \hbox{$w=2(A-B\cos\Theta)$.}
Since at zero momenta the total relativistic energies are nonzero, it is more physical to consider
the kinetic energy only, $K_i=E_i-m$. The rule for the kinetic energy composition is hence given by
\ba
 K_{12} &=& K_1 + K_2 + \frac{F(2A+2B)-F(2A-2B)}{4B} 
\nl
& & - U(2mK_1) - U(2mK_2) + U(0).
\ea{KIN_COMPOSE}
The coefficients $A$ and $B$ are also expressed by the respective kinetic energies:
\ba
 A &=& m(K_1+K_2)+K_1K_2, \nl
 B &=& K_1K_2\left(1+2m/K_1\right)^{1/2}\left(1+2m/K_2\right)^{1/2}. 
\ea{AB_KIN}
One observes that the product of kinetic energies occurs due to kinematic reasons.


Taylor expanding the integral of the unknown function $U(w)$ around $w=2A$
we obtain the following composition rule for the relativistic kinetic energies:
\ba
 h(x,y) &=& x + y - U(2mx) - U(2my) +U(0) \nl
 && + \sum_{j=0}^{\infty} U^{(2j)}(2A) \, \frac{(4B^2)^j}{(2j+1)!}
\ea{KINETIC_COMPOSITION}
with $A=m(x+y)+xy$ and $4B^2=4xy(x+2m)(y+2m)$.
The fiducial derivative becomes an expression with a finite number of terms
\ba
 h_2'(x,0) &=& 1 - 2m \, U'(0) + 2(m+x) \, U'(2mx) \nl
   && + \, \frac{4}{3} mx \, (2m+x) \, U''(2mx).
\ea{FIDUC_KINETIC}
For all traditional approaches the interaction energy $U$ is not considered as dependent
on $Q^2$. In such cases $h_2'(x,0)=1$ and one arrives at the simple addition as composition
rule. As a consequence the stationary energy distribution is of Boltzmann-Gibbs type.
For $Q^2$ dependent interaction it is enlightening to analyze  two particular kinematical cases: 
the extreme relativistic and the non-relativistic ones.
In the first case $m=0$ has to be replaced and one obtains
\be
 h_2'(x,0) = 1 + 2x \, U'(0).
\ee{FIDUC_EXTREME}
As discussed above this leads to a Tsallis-Pareto distribution in the kinetic energy
(at zero rapidity in the variable $m_T-m$).
The opposite extreme, $m \gg x$ leads to an undetermined asymptotic composition rule due to
\be
 h_2'(x,0) \approx 1 + 2m \left(U'(2mx)-U'(0)\right) + \frac{8}{3} m^2 x U''(2mx).
\ee{NONREL_FIDUC}
This result includes for $U'=0$ the traditional momentum independent interaction
case leading to the addition as asymptotic
rule for non-relativistic kinetic energies, and hence to the Boltzmann-Gibbs distribution.
We note that in the relativistic kinematics the next simplest assumption, 
$U'=\alpha = {\rm const.}$  leads to a 
Tsallis-Pareto distribution due to $h_2'(x,0)=1 + 2\alpha x$ form eq.(\ref{FIDUC_KINETIC}),
while its non-relativistic approximation form eq.(\ref{NONREL_FIDUC}) 
is still the Boltzmann-Gibbs exponential.

\section{Conclusion}
Summarizing we have proved that in the thermodynamical limit, by composing a finite amount
of an extensive physical quantity as a repeated composition of infinitesimally small amounts
of the same quantity one arrives at an effective asymptotic composition rule which is derivable
from a formal logarithm. This formal logarithm, $L$ is constructed from the original composition rule
uniquely and serves as a basis of describing a stationary distribution, $exp \circ L$ and the
corresponding formal expression for the entropy it canonically maximizes, 
$\langle L^{-1}\circ \ln\rangle$. Associative rules lead to themselves in this limit, and are 
attractors for other rules.

The addition is the simplest composition rule, the formal logarithm being the identity map.
It leads to the Boltzmann-Gibbs distribution.
The next simplest one leads to the Pareto-Tsallis power-law tailed stationary distribution.
We also have shown that considering interaction energies dependent on the relative four-momentum,
in general a nontrivial asymptotic rule arises. In particular the Pareto-Tsallis distribution
- and the corresponding non-additive asymptotic composition rule -
emerges generally from extreme relativistic kinematics. In fact high energy particle spectra
often show a power-law like tail in their kinetic energy. Besides that also non-relativistic systems
may show such thermodynamical limit, but here the general case can also be different.

\vspace{3mm}
{\bf Acknowledgments}

This work has been supported by the Hungarian National Science Fund, OTKA
(K49466, K68108). Discussions with
C.~Tsallis,  P.~V\'an, A.~L\'aszl\'o and Z.~R\'acz
are gratefully acknowledged.


\end{document}